\documentclass[preprint,nofootinbib,nobibnotes,superscriptaddress,longbibliography,floatfix]{revtex4-2}
\usepackage{amssymb}
\usepackage{graphicx,amsmath}
\usepackage{bm}
\usepackage{times}
\usepackage{epstopdf}
\usepackage{xcolor}
\def\expect#1{\left\langle #1 \right\rangle}

\newcommand{\edit}[1] {#1}

\begin{document}

\title{Integrated optically pumped magnetometer for measurements within Earth's magnetic field}
\date{\today}
\author{G. Oelsner}
\email{gregor.oelsner@leibniz-ipht.de}
\affiliation{Leibniz Institute of Photonic Technology, P.O. Box 100239, D-07702 Jena, Germany}
\author{R. IJsselsteijn}
\affiliation{Supracon AG, An der Lehmgrube 11, D-07751 Jena, Germany}
\author{T. Scholtes}
\affiliation{Leibniz Institute of Photonic Technology, P.O. Box 100239, D-07702 Jena, Germany}
\author{A. Kr\"uger}
\affiliation{Supracon AG, An der Lehmgrube 11, D-07751 Jena, Germany}
\author{V. Schultze}
\affiliation{Leibniz Institute of Photonic Technology, P.O. Box 100239, D-07702 Jena, Germany}
\author{G. Seyffert}
\affiliation{Optikron GmbH, L\"obstedter Str. 70, D-07749 Jena, Germany}
\author{G. Werner}
\affiliation{Fibotec Fiberoptics GmbH, Herpfer Str. 40, D-98617 Meiningen, Germany}
\author{M. J\"ager}
\affiliation{Supracon AG, An der Lehmgrube 11, D-07751 Jena, Germany}
\author{A. Chwala}
\affiliation{Leibniz Institute of Photonic Technology, P.O. Box 100239, D-07702 Jena, Germany}
\author{R. Stolz}
\affiliation{Leibniz Institute of Photonic Technology, P.O. Box 100239, D-07702 Jena, Germany}
%\pacs{42.50.Pq, 32.60.+i, 32.60.+i, 85.25.Cp, 75.45.+j}

\begin{abstract}
We present a portable optically pumped magnetometer instrument for ultra-sensitive measurements within the Earth's magnetic field.
The central part of the system is a sensor head operating a MEMS-based Cs vapor cell in the light-shift dispersed $M_z$ mode.
It is connected to a compact, battery-driven electronics module by a flexible cable.
We briefly review the working principles of the device and detail on the realization of both, sensor head and electronics.
We show shielded and unshielded measurements within a static magnetic field amplitude of 50~$\mu$T demonstrating a noise level of the sensor system down to 140~fT/$\sqrt{\textrm{Hz}}$ and a sensor bandwidth of several 100 Hz.
In a detailed analysis of sensor noise, we reveal the system to be limited by technical sources with straightforward strategies for further improvement towards its fundamental noise limit of 12~fT/$\sqrt{\textrm{Hz}}$.
\edit{We assess the parameters defining the sensor bandwidth by theoretical modeling based on the Bloch equations.}
Finally, we compare our sensors' performance to a commercial SQUID system in a  measurement environment typical for geomagnetic observatory practice and geomagnetic prospection.
\end{abstract}
\maketitle

\section{Introduction}
The ultra-sensitive evaluation of magnetic fields is an important tool throughout various research areas and applications.
\edit{For example in medicine} biomagnetic signals of the brain \cite{Hmlinen1993} or the heart \cite{Koch2004} recorded in magnetically well-shielded rooms can help diagnostics.
Other important applications lay in the fields of geophysical exploration, geomagnetism, and archeology, where magnetic measurements at highest sensitivity are required within Earth's magnetic field strength \cite{Clark1997,Nabighian2005}.
To date, in most of these applications demanding for ultimate performance, superconducting quantum interference devices (SQUIDs) are the sensors of choice \cite{Stolz2001,Foley2004,Stolz2015}.
However, recent developments in the field of optically pumped magnetometers (OPMs) \cite{Budker2013}
pave the way towards OPM systems competitive with SQUIDs with respect to magnetic field resolution, but featuring additional advantages, e.~g., avoiding the need for cryogenics or gaining higher flexibility due to smaller sensor sizes \cite{Shah2007,Boto2018}.

To date, the best sensitivities of OPMs are achieved using the so-called spin-exchange relaxation-free (SERF) mode \cite{Allred2002,Kominis2003,Dang2010}.
Still, SERF requires operation in nearly zero magnetic field and strongly limits the signal bandwidth.
Thus, relying on heavily magnetically shielded rooms and low-frequency measurements, SERF OPM application scenarios are limited and their use within the Earth's magnetic field is not feasible.
OPMs developed for the use at medium-sized magnetic field amplitudes (few to tens of $\mu$T) show sensitivities down to tens of fT \cite{Aleksandrov2009} or even sub-fT as demonstrated in the lab \cite{sheng2013} using large glass-blown vapor cells.
However, because of the large cell volumes such sensors typically suffer from increased vulnerability to magnetic field gradients when used outside the lab and offer only a rather limited sensor bandwidth.
Recently, several groups have presented OPM lab implementations aimed at further miniaturization using micro-fabricated vapor cells and integrated optics packaging \cite{guo2019,zhang2019} or showing methods to obtain noise levels on the order of 100~fT/$\sqrt{\textrm{Hz}}$ in $\mu$T fields or even at Earth's magnetic field strength using gradiometric approaches and/or pump-probe schemes, e.g., \cite{lucivero2014,bevilacqua2016,deans2018,zhang2020,Limes2020}.

In Table~\ref{Tab:comm} we have compiled an overview on currently commercially available OPMs developed for the use in the Earth's magnetic field and their \edit{noise floor as well as sensor bandwidth} given in their respective data sheets.
While the magnetic field noise spectral density of the most sensitive commercial devices is 200~fT/$\sqrt{\textrm{Hz}}$, these offer only very limited bandwidth. On the other hand, other OPM types with increased sensor bandwidth show sensitivities in the pT-range.

\begin{table}[htb]
\begin{ruledtabular}
\begin{tabular}{l l c c}
company & sensor & noise level & bandwidth   \\
 &  & (pT/$\sqrt{\textrm{Hz}}$) & (Hz)   \\
 \hline\noalign{\smallskip}
GEM systems\footnote{https://www.gemsys.ca} & GSMP-35 & 0.2$^\bullet$ & 10 \\
GEM systems & GSMP-25 & 22$^\bullet$ & 10 \\
Geometrics\footnote{https://www.geometrics.com} & G-864  & 4 & 5$^\ast$  \\
Geometrics & MFAM\cite{Zhang2016}  & $1$ & 400  \\
QuSpin\footnote{https://quspin.com} & QTFM & $<1$ & 200$^\ast$ \\
Scintrex\footnote{https://scintrexltd.com} & CB-3 & 25 & 1$^\diamond$ \\
Scintrex & CS-3 & 0.6 & 0.1$^\diamond$ \\
Twinleaf\footnote{https://twinleaf.com}  & microSAM  & 20 & 100 \\
Twinleaf  & OMG  & $<0.2^\clubsuit$ & 1000 \\
\end{tabular}
\end{ruledtabular}
\caption{List of commercially available OPMs for the use within Earth's magnetic field. \edit{The noise level and bandwidth parameters are gathered from datasheets and information available on the company websites}. $\!^\bullet$ Value is given at 1~Hz. $\!^\ast$ Derived as the half of the specified maximal sampling rate. $\!^\diamond$ Maximal bandwidth for which sensitivity is given in the data sheets. \edit{$^\clubsuit$ The values for Twinleaf OMG are valid for magnetometer operation according to a personal note by T.~Kornack. We note that commercial systems still have advantages over the demonstrator system presented here in other key parameters like sensor footprint and power consumption.}}\label{Tab:comm}
\end{table}

Our work is based on OPMs operating in the light shift-dispersed $M_\textrm{z}$ mode (LSD-Mz) \cite{Schultze2017}.
We have employed a micro-fabricated integrated array of alkali vapor cells \cite{Woetzel2011} for measurements within finite magnetic fields and revealed the potential of sensitivities down to tens of fT/$\sqrt{\textrm{Hz}}$ at Earth's magnetic field strength with a sensor bandwidth approaching the kHz range \cite{Schultze2017}, overcoming main limitations of SERF magnetometers.
Recently, we have studied theoretically and experimentally the main characteristics of LSD-Mz, including the dependence of the sensor's reading (i.e. the so-called heading error) \cite{Oelsner2019} and sensitivity (i.e. the so-called dead zones) \cite{Oelsner2019a} on the orientation within the magnetic field when used at Earth's magnetic field strength.
In this paper we present the implementation of a complete sensor system based on LSD-Mz into an easy-to-use and robust measurement device.
Such a translational process is connected to a variety of technical challenges which must be identified and solved.
Moreover, the performance of a sensor system needs to be evaluated as a result of the interplay of possible fundamental sensor limitations, technical noise sources arising with the sensor integration as well as the testing conditions.
Thus, we experimentally investigate and discuss the functionality of our device in shielded and unshielded (within Earth's magnetic field) environment.

The paper is organized as follows: In Sec. \ref{Sec:WorkPrinc} we present a short description of the operation principles.
In Sec. \ref{Sec:system} we detail on the realization of the sensor module, i.~e., the integrated vapor cell, built-in optics, and the thermal design, as well as the compact electronics system.
After discussing the performance of our device in terms of sensor bandwidth and noise level in shielded and unshielded environment in Sec. \ref{Sec:perf}, we conclude the paper with a summary and an outlook.
\section{Working principle}
\label{Sec:WorkPrinc}
OPMs are based on the Zeeman effect, that is the shift of atomic energy levels in the presence of a static magnetic field \cite{Zeeman1897}.
Commonly alkali metal vapors are used as a sensor medium, due to their unique electron configuration featuring a single unpaired valence electron that is highly sensitive to external magnetic fields while convenient to manipulate by optical means.
Transitions between electronic ground and excited states in these alkali vapors are accessible by light in the visible or near-infrared range enabling efficient creation of atomic polarization by optical pumping \cite{Happer1972}.
In our sensor, we exploit the Zeeman splitting in the $F=4$ hyperfine level of the ${}^{133}$Cs atomic ground state as a function of the magnetic field $\nu_\mathrm{L} = \gamma |\vec{B}|$, with $\nu_\mathrm{L}$ the Larmor frequency and $\gamma \simeq 3.5$ kHz/$\mu$T the gyromagnetic ratio.
In order to measure $\nu_\mathrm{L}$, we employ optically detected magnetic resonance (ODMR) using a laser resonant with the Cs $D_1$ ($6S_{1/2}\rightarrow6P_{1/2}$) transition at a wavelength $\lambda\approx 895$~nm for both optical pumping and signal read-out.

\begin{figure}[htb]
  \includegraphics[width=7 cm]{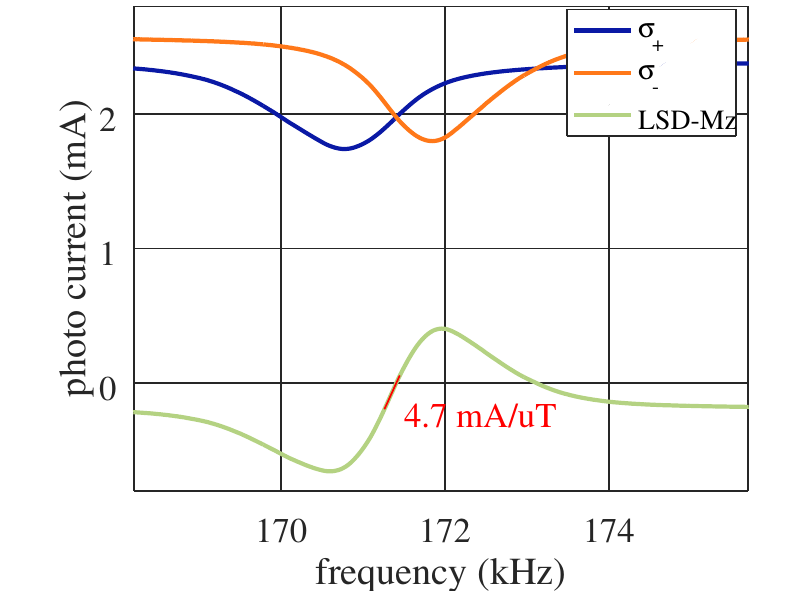}
  \caption{Measured output signals on the AD-converters of the two channels as a function of the applied rf-frequency $\nu_\mathrm{rf}$ in unshielded environment. The signals are converted from currents to voltages by transimpedance amplifiers here with a gain of $G = 10^3$~V/A. The channels of opposite helicity show resonant Lorentzian curves separated due to the dynamical AC Stark effect introduced by the strong off-resonant optical pumping. The difference signal features a steep linear transfer factor $s$ at the unshifted Larmor frequency $\nu_\mathrm{L}$. We discuss the influence of the different resonance amplitudes in the main text.} \label{Fig:signals}
\end{figure}

The LSD-Mz-mode \cite{Schultze2017} is a modified and extended $M_z$ magnetometer scheme \cite{Aleksandrov2009}.
The dc absorption / transmission of a circularly-polarized laser beam propagating in parallel to the magnetic field vector is measured as a function of the oscillation frequency $\nu_\mathrm{rf}$ of an applied radio-frequency (rf) magnetic field.
\edit{Therefore, the sensors suffers from the same angular dependence of sensitivity on sensor orientation with respect to external magnetic field direction as known from the traditional $M_z$ magnetometers, as demonstrated in Ref.~\onlinecite{Oelsner2019a}.
In comparison, it additionally exploits the light narrowing effect \cite{Scholtes2011,*Scholtes2011e}, namely large spin polarization by simultaneous pumping of both hyperfine ground states and uses two magnetic resonance that are detuned by large dynamical ac Stark shifts $\Delta$ using laser beams with opposite circularly polarized laser beams. }
As the magnitude and sign of $\Delta$ depends on the intensity as well as the degree and handed-ness of the circular polarization of the laser beams ($\sigma_\pm$: $\Delta\sim \mp1$~kHz), the magnetic resonance frequencies in the two channels are (symmetrically) shifted away from each other assuming perfectly matched conditions.
When the ac Stark shift is on the order of the magnetic resonance linewidth, taking the difference of the $M_z$ signals in the two channels results in a dispersive curve with a quasi-linear part and a large steepness (or transfer factor) $s$ at the unshifted Larmor frequency $\nu_\mathrm{L}$ as shown together with the individual Lorentzian-shaped $M_z$ signals in Fig.~\ref{Fig:signals}.
The dispersive LSD-Mz signal is well-suited for sensor operation and generated without the need for elaborate phase-sensitive (lock-in) detection, which might be expensive and prone to systematic shifts due to phase errors or instabilities in the electronic detection circuit.
The excellent single channel signal contrast and large steepness $s$ allows the LSD-Mz magnetometer to reach down to a fundamental noise limit in the 10~fT/$\sqrt{\textrm{Hz}}$ range using vapor cells of few mm size \cite{Schultze2017,Oelsner2019a} (see Sec.~\ref{Sec:sensi} for details).
In addition, because of the differential nature of the LSD-Mz signal, sources of noise common to both channels, e.g., excess technical laser intensity, polarization noise, and fluctuations, are intrinsically suppressed in the case of well-balanced channels.
In Fig.~\ref{Fig:signals} the two signal curves show slightly different amplitudes and offsets.
This indicates that the total intensity, grade of helicity, profiles of the laser beams, or the local amplitude of the rf-field are not equal at the cell positions.
Thus, as the zero crossing of the dispersive signal is shifted, the accuracy of the sensor is limited.
In addition, at this frequency point the steepness of the individual signals is accordingly different.
While this has a very limited effect on the sensor sensitivity, it leads to imperfect compensation of common noise in the difference signal.
This problem can, however, be alleviated by introducing a scaling factor to one of the channels in post processing.
Nevertheless, the experimental curves illustrate the crucial importance of balancing the two beams' intensities, powers, and helicities for optimal sensor performance.

\edit{
The magnetic resonances shown in Fig.~\ref{Fig:signals} have a width of about 1~kHz that is common for the parameters used in the LSD-Mz operational mode.
This property allows to expect large signal bandwidths because for conventional rf-driven OPMs without feedback loop it is given by the width of the employed magnetic resonance signal, i.e., the frequency response of the OPM corresponds to a low-pass filter with a cut-off frequency given by the magnetic resonance width.
However, in the LSD-Mz method an additional peculiarity arises as it operates with an rf frequency signal which is detuned by $\pm \delta$ from the Mz magnetic resonance points of the two $\sigma_{\pm}$ channels.
To estimate the signal bandwidth as a first step, we write the Hamiltonian of one channel in two-level approximation \cite{Oelsner2019a} and extend it by an oscillating magnetic test signal with frequency $\nu_s$ and an amplitude $\Omega_s$ in frequency units
\begin{equation}
H = h\frac{\nu_\mathrm{L} + \Omega_s \cos 2\pi \nu_s t}{2} \sigma_z +h \Omega_{rf} \cos 2\pi \nu_\mathrm{rf} \sigma_x,
\end{equation}
where we have additionally introduced the amplitude $\Omega_{rf}$ of the rf-field, the Pauli operators $\sigma_i$ as well as Planck's constant $h$.
We include dissipative dynamics given by the rates $\Gamma_r$ and $\Gamma_\varphi$ corresponding to respective inverse $T_1$ and $T_2$ times, and solve the resulting Bloch equations in the rotating frame, similarly as, e.g., in Refs.~\cite{Shevchenko2014,Oelsner2020}.
Because of the non-linearity of the vapor system, the solution for the populations' expectation value that corresponds to the magnetization in z-direction might be expressed as sum over harmonics of the signal tone
\begin{equation}
\expect{\sigma_z} = \expect{\sigma_z}_0 + \sum_{m=1}^\infty \left( A_m e^{im\omega_s t} + B_m e^{-im\omega_s t}\right).
\end{equation}
Expanding the Bloch equations to first order in $\Omega_s/\nu_s$ allows to express the oscillation of the population at the signal frequency as
\begin{equation}\label{Eq:output}
\expect{\sigma_z}_1 = s_1 \cos \left(2\pi \nu_s t + \alpha \right).
\end{equation}
In the optical detection scheme this oscillation of population is translated into a modulation of the output photo current at the signal frequency $\nu_s$.
Its amplitude might be found by multiplying $s_1$ with the photo current away from the magnetic resonance \cite{Oelsner2019a}.
The amplitude $s_1$ and phase $\alpha$ of \eqref{Eq:output} are given by
\begin{equation}
\begin{split}\label{Eq:amp_and_phase}
s_1 &= \Omega_s \frac{\Omega_{rf}^2 }{2 \nu_s} \expect{\sigma_z}_0^2 \text{sgn}(\delta) \sqrt{\frac{I^2 + R^2}{\Gamma_r^2 + \nu_s^2 \expect{\sigma_z}_0^2}}, \\
\alpha &= \left(\Gamma_r I + \expect{\sigma_z}_0 \nu_s R \right)/\left(\Gamma_r R - \expect{\sigma_z}_0 \nu_s I \right),
\end{split}
\end{equation}
where $\text{sgn}$ denotes the sign function and we have introduced the abbreviations
\begin{equation}
\begin{split}
I &= \Gamma_\varphi \left( \frac{1}{\left( \delta + \nu_s\right)^2 + \Gamma_\varphi^2} -\frac{1}{\left( \delta - \nu_s\right)^2 + \Gamma_\varphi^2}\right), \\
R &= \frac{2\delta}{\delta ^2 + \Gamma_\varphi^2} -\frac{\delta-\nu_s}{\left( \delta - \nu_s\right)^2 + \Gamma_\varphi^2} -\frac{\delta + \nu_s}{\left( \delta + \nu_s\right)^2 + \Gamma_\varphi^2},\\
\expect{\sigma_z}_0 &= -\frac{\Gamma_r \Gamma_\varphi^\prime}{ \Gamma_r \Gamma_\varphi^\prime +  \Omega_{rf}^2}, \quad \Gamma_\varphi^\prime = \frac{\delta^2+\Gamma_\varphi^2}{\Gamma_\varphi}.
\end{split}
\end{equation}
The LSD-Mz signal is given as the difference of two signals \eqref{Eq:output} shifted from each other by the ac Stark shift for $\delta = \pm \Delta$ with the working point of the sensor given at the crossing of the two magnetic resonance signals.
As required for a sensor, the output signal amplitude is proportional to the signal amplitude $\Omega_s$.
The transfer factor depends on detuning, driving amplitude, dissipative rates, as well as on the signal frequency itself.
Moreover, in the limit of small signal frequencies, the transfer factor is given by the first derivative of the steady state expectation value $\expect{\sigma_z}_0$ for $\Omega_s = 0$ representing the steepness $s$ as a function of the detuning $\delta = \nu_L-\nu_{rf}$.
The output signal amplitude reduces with increasing values of $\nu_s$.
The cut-off frequency or signal bandwidth is determined by the longitudinal and transverse relaxation times, $T_1$ and $T_2$ of the spin polarization of the alkali vapor atoms.
In addition to that, resonances of the signal frequency with the detuning $\delta = \pm\nu_s$ can create local maxima in the amplitude vs. signal frequency plots.
This can enable a slight extension of the bandwidth.
The phase factor $\alpha$ strongly depends on the signal frequency at these resonance points.
For an asymmetric detuning of the $\sigma_+$ and $\sigma_-$ channels a small phase offset in the output signals can be expected.
However, in our system, such phase offset has only a marginal influence, because the LSD-Mz signal amplitude found as by subtracting the two channels has only a cosine dependence on this phase difference.
}
\section{The LSD-Mz magnetometer system}
\label{Sec:system}
The LSD-Mz magnetometer system presented here consists of a sensor module and the corresponding supply and data read-out electronics box.
The sensor module includes the alkali vapor cell structure, an integrated optical setup for beam preparation and detection as well as the coil to generate the rf field driving the atomic magnetic resonance.
%
%Miniaturized vapor cells \cite{Woetzel2011} are employed to avoid broadening of the magnetic resonance due to possible magnetic field gradients appearing in field use.
%
%Thus, in order to achieve an adequate optical density, the alkali vapor cell structure has to be heated.
%
%Main tasks of
\edit{The electronics box ensures a reliable operation of the used laser diodes and rf-field generations, enables amplification and digitization of the photo diode signals, as well as calculation of the differential sensor signal and deployment of an interface for data recording.
%
%Electronic feedback circuits are required for temperature control of the laser diodes and the alkali vapor cell, as well as for tracking of the Larmor frequency by the rf field frequency in the locked sensor mode.
%
The magnetometer system is designed for field experiments and thus laid out to be compact, robust, and battery-driven.}

\subsection{The sensor module}
\label{Sec:Module}
\begin{figure}[htb]
  \includegraphics[width=8 cm]{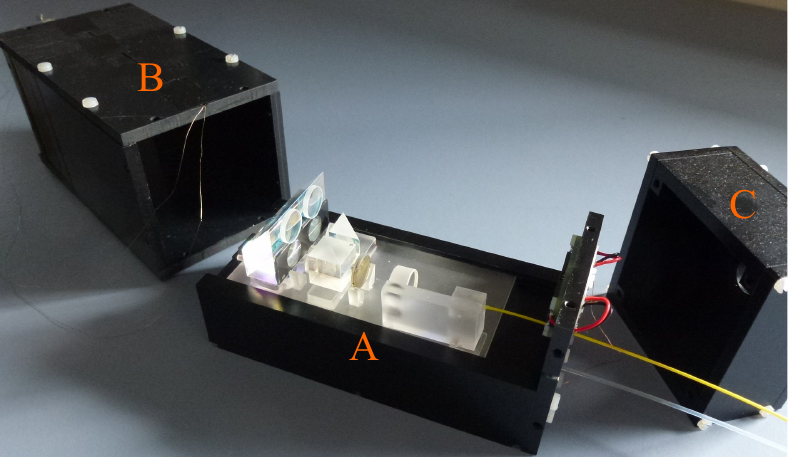}
  \caption{Photo of the sensor module. It consists of three parts: Part A holds the main optical components and the vapor cell. Part B serves as outer housing for the optics and the vapor cell and encloses the laser radiation. It features additionally the rf-coil for the magnetic resonance. In part C the electrical wires and optical fibers are supported against tension and these lines are bundled together for the connection to the electronics.}\label{Fig:module}
\end{figure}

\begin{figure}[htb]
  \includegraphics[width=8 cm]{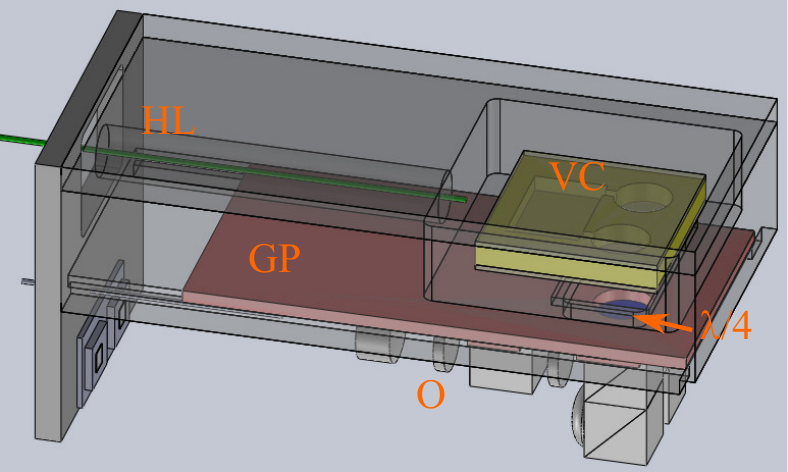}
  \caption{Concept sketch of the sensor module's central part (top down view of A in Fig.~\ref{Fig:module}). For better visibility the heating laser's fiber (HL), the vapor cell (VC), the supporting glass plate (GP) of the optics (O), and the quarter lambda plates ($\lambda$/4) are colored separately in green, yellow, red, and blue, respectively. }\label{Fig:module2}
\end{figure}

The main functionality of the OPM is implemented into a sensor head with outer dimensions of 69$\times$69$\times$167 mm$^3$.
As shown in the photograph image of Fig.~\ref{Fig:module} it can be disassembled into three parts.
The whole construction is made from non-magnetic polyoxymethylene (POM) with a minimal wall thickness of 5~mm.
In the concept view of Fig.~\ref{Fig:module2} the main components of the central part A are visible.
The optical elements for pump beam preparation (O) are mounted on an additional glass plate (GP) for accurate alignment and mechanical stability.
The vapor cell (VC) is fixed in a special compartment of the module and heated by laser light guided to the cell by a multi-mode fiber (HL).
A detailed discussion of these parts is given below.
The outside box (B in Fig.~\ref{Fig:module}) is detachable and features the coil used to apply the rf-field needed for ODMR.
The latter has a quadratic base of 65$\times$65~mm$^2$ and a length of 65~mm.
The wire is placed in milled grooves in distances of 2~mm containing two windings each.
Additionally, in the last three notches on each side, the number of turns is doubled to four to improve the field homogeneity produced at the vapor cells positions.
For a superior stability of the electrical resistance with temperature and to reduce the influence of the self resonance of the coil we have chosen a manganin wire over copper.
\subsubsection{Cs vapor cell}
\label{SSec:cell}
\begin{figure}[htb]
  \includegraphics[width=3.5 cm]{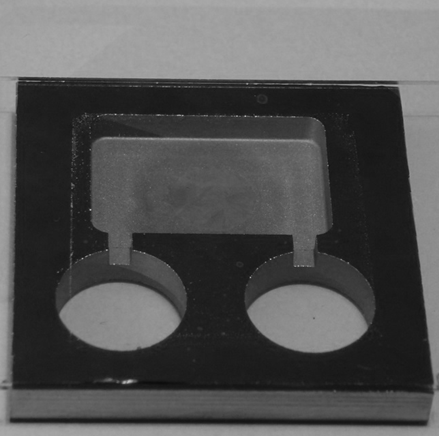}
  \caption{Photograph of the used vapor cell. Two identical microfabricated alkali vapor cells connected to a shared Cs reservoir form the sensing volume of the sensor module. The outer dimensions of the structure are 35x35x6~mm$^3$ while the radius of the sensing cavities is 5.5~mm and their centers are 16~mm apart.}\label{Fig:cell}
\end{figure}

The central element of the sensor module is the vapor cell structure as shown in Fig.~\ref{Fig:cell}.
It is created by microfabrication technologies.
In a first step, the basic geometry of the cells and the Cs reservoir is structured into a 4~mm thick silicon wafer by ultrasonic milling.
After cleaning \edit{the structured silicon wafer}, a glass plate is attached at the bottom of the Si wafer by anodic bonding.
The glass features an anti-reflection (AR) coating for the pumping wavelength of 895~nm on the inside and a dielectric mirror at the outside of the cell. The latter is used for the reflection of the pumping beams on the bottom of the cell.
The cell is filled with droplets of diluted cesium azide.
After drying for about one hour at RT as well as two hours at $200^{\circ}\mathrm{C}$ in vacuum a second glass plate with AR coating layers on top and bottom is mounted to the top of the cell by anodic bonding, hermetically sealing the cell.
In a final step, the cesium azide is decomposed by irradiation with an excimer laser into pure cesium metal and nitrogen.
The latter is used as buffer gas. A detailed description of the cell fabrication process can be found in Ref.~\onlinecite{Woetzel2011}.
Due to a buffer gas pressure of about 200~mbar at RT, the optical transition lines of the Cs D$_1$ line are broadened to a full width at half maximum (FWHM) of about 4~GHz.
A typical optical spectrum of such a buffer gas cell can be found in Ref.~\onlinecite{Scholtes2011,*Scholtes2011e} and the influence of the buffer gas pressure is, for example, discussed in Ref.~\onlinecite{Schultze2015}.
\subsubsection{Beam preparation}
\begin{figure}[htb]
  \includegraphics[width=8 cm]{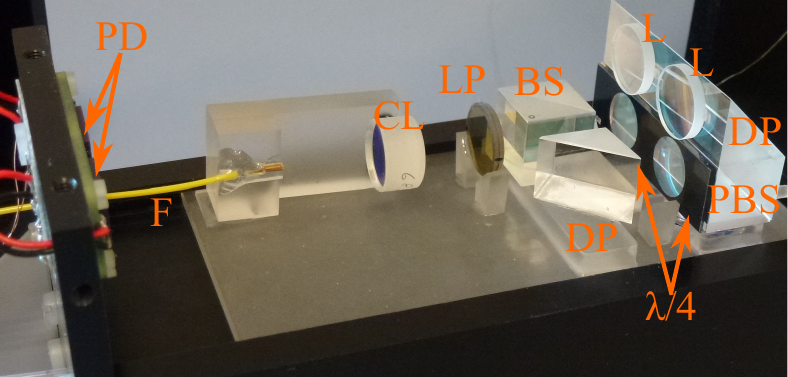}
  \caption{Photograph of the optical setup included in the sensor head for preparation of the pump beams and their detection. A detailed description is given in the main text. The optical components have a characteristic size of 12 mm. }\label{Fig:optics}
\end{figure}

Laser light at a wavelength of 895~nm is generated by a distributed Bragg reflector (DBR) laser diode and guided to the sensor module by a ten meter long single-mode polarization-maintaining fiber (F).
At each of the vapor cells a laser power of roughly 2~mW per beam is required.
The optics on the sensor head, as shown in Fig.~\ref{Fig:optics}, are used for the preparation of the pump beams and their subsequent detection.
A collimating lens (CL) is used for stable collimation of the beam to a diameter of about 4~mm.
This beam further is passed through a linear polarizer (LP) that converts polarization noise into amplitude noise and defines the direction of the lights' linear polarization axis.
By the use of a 1:1 beam splitter (BS) and a deflection prism (DP) two parallel beams are created.
These beams are deflected by polarizing beam splitters (PBS) and acquire circular polarization by two separately adjustable quarter-wave plates ($\lambda/4$, see also Fig.~\ref{Fig:module2}).
After passing the magnetometer cell twice (reflected at the lower vapor cell glass layer by the dielectric mirror), the beams pass through the same quarter-wave plates a second time, acquiring a linearly polarized state with the axis rotated by $\pi/2$ compared to the incoming light, thus being transmitted straight through the PBS.
With the use of additional DPs and lenses (L), the beams are then slightly focused onto two photodiodes (PD), where the total light power is transformed into a proportional photo current.

The whole optical setup is mounted on a 2~mm thick glass plate.
The quarter-wave plates as well as the linear polarizer are aligned for optimal balancing of the created laser beams featuring opposite helicity with respect to an equal degree of circular polarization as well as an equal power.
\edit{All optical components are glued in place using NOA 61.}
\subsubsection{Thermal regime}
For a sufficient optical density of the Cs vapor employed in the vapor cell (cf. Fig.~\ref{Fig:cell}), it is required to heat the cell to about 100$^{\circ}$C.
To do so, the cell is irradiated at one of its side walls by an infrared diode laser with a wavelength of 808~nm.
This laser light is guided to the sensor module by a multi-mode fiber as can be seen in Fig.~\ref{Fig:module2}.
On the irradiated side of the vapor cell a soot layer is deposited for optimal absorption of the heating laser that has an approximate spot size of about 2~mm.
The large heat conductivity of the silicon ensures an almost homogenous temperature distribution across the cell structure while the glass surface at the windows and especially at the reservoir is slightly colder, see Ref.~\cite{IJsselsteijn2012}.
A cold finger is placed on the Cs reservoirs' top glass layer to define the coldest spot of the vapor cell, thus avoiding condensation of Cs metal on the optical windows.
A platinum metal resistor (Pt100) glued to the vapor cell is used for temperature monitoring and control.
\subsection{Supply and data read-out electronics}
\label{Sec:electronics}
\begin{figure}[htb]
  \includegraphics[width=8 cm]{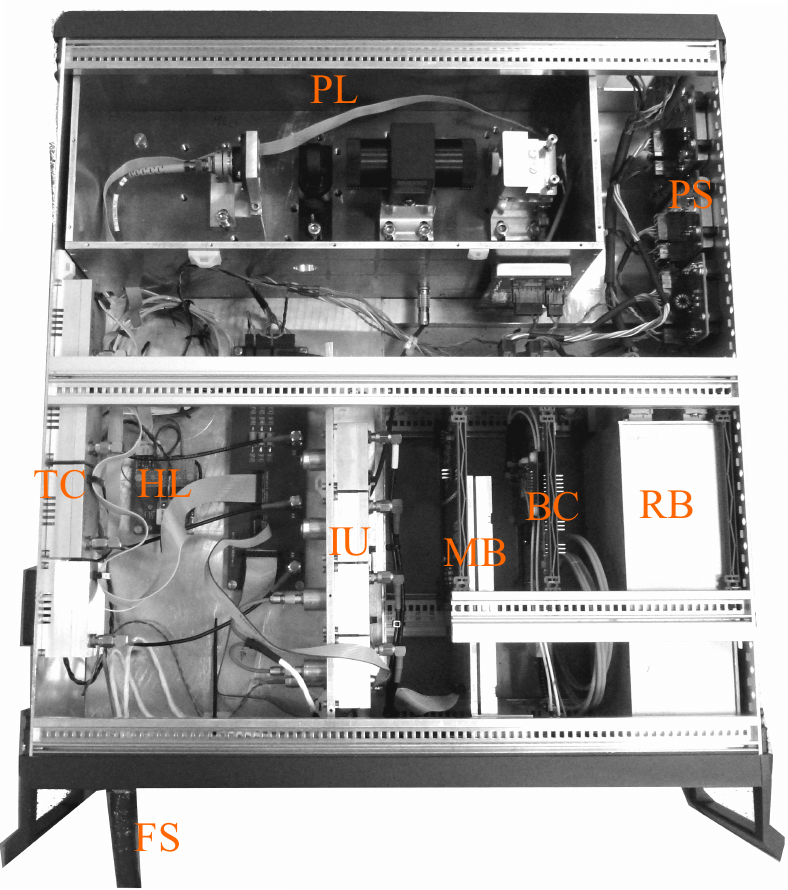}
  \caption{Photo of the control electronics. The different building blocks, namely pumping laser (PL), power supply (PS), temperature controller (TC), heating laser (HL), transimpedance amplifiers (IU), main board including a FPGA (MB), bus controller (BC), and rechargeable batteries (RB), are separated in different compartments. The cables and fibers connecting the sensor module to the electronics are protected against external disturbances by the use of a fire sleeve (FS). A detailed description is given in the main text.}\label{Fig:electronics}
\end{figure}

The sensor module is connected to the supply and data read-out electronics by a ten meter long fire sleeve containing the electrical cables for the currents generated at the photo diodes, temperature readout, and rf-coil supply as well as the heating and pumping laser fibers.
The sleeve ensures protection from mechanical stress and environmental disturbances.
A photograph of the electronics is shown in Fig.~\ref{Fig:electronics}.
The electronics module is integrated into a 19'' rack with a total weight of about 17~kg.
The electronics is based on a field programmable gate array (FPGA) board for signal control, fast real-time calculations, and signal routing.
A bus controller connected by a fiber interface enables the communication with a standard laptop computer for control and data acquisition via USB.
The FPGA controls the laser current of the DBR laser that takes a value of about 200~mA.
The laser current source features a noise of 1~nA/$\sqrt{\text{Hz}}$ @ 1~kHz.
Additionally, a low-pass filter is included for suppressing high-frequency current noise.
The DBR laser diode is included into a separately closed compartment and features a spectral linewidth $<2$~MHz.
A temperature controller is used for stabilization of the laser temperature to better than 2~mK using a Peltier cooler and a negative temperature coefficient (NTC) thermistor for monitoring.
Both are included in the commercial DBR laser diode package.

By adjusting the laser diode temperature the laser light frequency can be tuned at a rate of 22~MHz/mK.
During measurements the laser frequency is not actively stabilized but initially set to maximize the LSD-Mz signal slope.
The maximum light power delivered by the DBR diode is roughly 240~mW and can be adjusted with the aid of neutral density (ND) filters inserted into the free beam.
An adjustable lens integrated onto the laser diode package allows collimating the laser beam.
The beam is passed through an optical isolator before coupled into the fiber to avoid optical feedback on the laser diode.
Our investigations show that mechanical vibrations affecting the fiber coupling are to a large extent responsible for low-frequency sensor noise.
%%%%

A similar supply and control is used for the fiber-coupled single emitter diode laser used for heating of the alkali vapor cell.
Its power can be adjusted by the laser current to a maximum value of 7~W.
A temperature regulation to $20^{\circ}\mathrm{C}$ is used to protect the laser diode.
The wavelength of the heating laser at $\lambda=808$~nm is chosen to be far off-resonant with the Cs absorption lines.
Therefore, its wavelength stability is not of critical importance.
The vapor cell temperature is regulated by another temperature control unit, which monitors the cell temperature by a Pt100 mounted to the vapor cell and controls the heating power by adjustment of the heating laser current.

Together with the two fibers, four ten meter long shielded twisted-pair electrical cables connect the sensor module to the electronics: One cable is used for the readout of the cell's PT100 and a second to apply current to the rf coil.
The rf frequency is created by direct digital synthesis (DDS) at a frequency and voltage resolution of 28 and 14~bit, respectively, in a frequency range up to 10~MHz.
Thus, the rf frequency can be set in steps of roughly 37~mHz, corresponding to about 10~pT in magnetic field units.
Directly in the supply unit, the generated voltage is converted into a current, the amplitude of which is adjusted for optimal width and amplitude of the magnetic resonances to a value of about 1~mA.

The two remaining electrical cables connect the photo diodes of the sensor head with the respective transimpedance amplifiers.
Their gain $G$ as well as the applied photo diode bias voltage are controlled via the FPGA.
The resulting voltage signals are finally converted by analog to digital converters (ADC) at 24 bit resolution, a sampling rate of $sr=32$~kHz, and a signal to noise ratio (SNR) of 113.5~dB.
The FPGA calculates the difference signal of the two signals which is eventually used for the frequency feedback control of the rf field.
A software-implemented proportional-integral (PI) controller is used to regulate the signal amplitude of the difference signal to zero in closed loop operational mode.
The value of this control signal is recorded in addition to the photo diode signals and difference signal, while the latter represents the remaining error signal.
\edit{The system is powered by two rechargeable batteries each having 16~V and a maximal charge of 10~Ah.
Most of the electrical power is drained by the heating laser requiring about 20~W.
By switching to electrical heaters the power required for cell heating can be reduced roughly by a factor of 5.
In the current state, the system can run for about 6 hours on battery power.
}

\section{Instrument performance}
\label{Sec:perf}
In the following we discuss results of measurements performed with the instrument in a magnetically shielded barrel (MSB) described in detail in Ref.~\cite{Schultze2010}, as well as in unshielded environment, viz., within Earth's magnetic field.
\begin{figure}[htb]
  \includegraphics[width=8 cm]{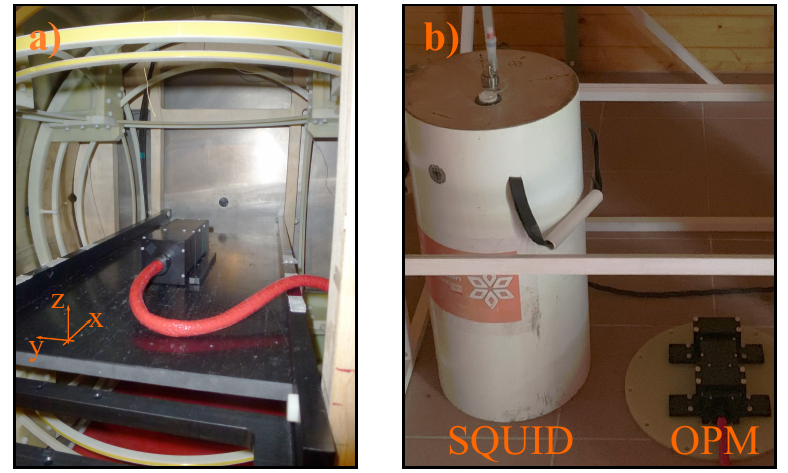}
  \caption{Measurement environments. a) For characterization of the sensors overall noise characteristic, it is placed inside of a MSB. b) The OPM sensor module is placed next to the cryostat of a three-axis SQUID system inside of a wooden hut without magnetic shielding. The control electronics for both systems (not shown) are placed in a distance of about 10 m.}\label{Fig:setups}
\end{figure}
\subsection{Sensor sensitivity and noise analysis}
\label{Sec:sensi}
For an in-depth sensor noise analysis we sort potential contributions limiting the noise floor into three categories: 1) fundamental limits connected to the OPM operation mode, 2) contributions from the sensor electronics as well as 3) limits set by the measurement environment.
\subsubsection{Fundamental sensitivity}
The quantum-mechanical uncertainty in the determination of the spin projection of an atomic ensemble sets a fundamental limit to the noise of quantum sensors.
In case of OPMs, the spin projection noise (or sometimes called the standard quantum limit, SQL) is given by \cite{Auzinsh2004,Aleksandrov2009}
\begin{equation}\label{Eq:spinprojlimit}
B_{sp} \simeq \frac{1}{2\pi\gamma} \sqrt{\frac{\Gamma}{N}}\,,
\end{equation}
where $N$ is the number of atoms within the active volume and $\Gamma$ represents the transverse relaxation rate in units of Hz, which can be identified with the full linewidth of the magnetic resonance signal.
For our system we estimate $B_{sp}$ from the number of atoms inside the volume defined by the beam\footnote{Due to high buffer gas pressure we neglect diffusive motion of atoms in and out of the beam volume.}
 $V_b = \pi r^2 h$ found from the ideal gas law
\begin{equation}
N = \frac{p V_b}{k_B T},
\end{equation}
with $r=2$~mm the radius of the beam, $h=4$~mm the length of the cylindrical vapor cell volume (which is passed twice), $T=100^{\circ}\mathrm{C}$ the cell temperature and $k_B$ the Boltzman constant.
The saturated vapor pressure of Cs vapor in N/m$^2$ is given by \cite{Alcock1984} as
\begin{equation}
p = 10^{5.006+A+B/T},
\end{equation}
with $A = 4.165$ and $B=-3830$~K for liquid Cs.
Assuming a transverse relaxation rate of 2~kHz \cite{Oelsner2019a} and considering the doubled number of atoms due to the use of two beams in the LSD-Mz mode, we find $B_{sp} = 11.4$~fT/$\sqrt{\text{Hz}}$.

Another typical fundamental limit for OPMs is the shot-noise arising from the discrete nature of the probing light.
It results from the quantum uncertainty of the number of photons inside of a laser beam.
For the LSD-Mz magnetometer the associated spectral magnetic noise density is
\begin{equation}\label{Eq:shotnoiselimit}
B_{sn} = \frac{1}{\gamma}\frac{\sqrt{2e \left(I_1+I_2\right)}}{s},
\end{equation}
which is a function of the measured mean photo current levels $I_i$ detected by the photo diodes of the two channels\footnote{assuming perfect conversion efficiency} and the steepness $s$ of the difference signal.

For our system we obtain $B_{sn} = 4.2~\mathrm{fT/\sqrt{Hz}}$ using $I_i=0.42$~mA and $s=3.92~\mathrm{mA/\mu T}$, which is about a factor of three smaller than $B_{sp}$.
\subsubsection{Electronics noise}
We expect possible noise contributions connected to signal digitization, and consequently estimate the noise spectral density ($NSD$) of the ADCs.
We start from the equation of the voltage noise spectral amplitude $\tilde{V}$ as
\begin{equation}
\tilde{V} = \frac{V_{rms}}{\sqrt{\Delta f}},
\end{equation}
with $V_{rms}$ the root mean square voltage noise and $\Delta f$ the bandwidth.
Calculating $20 \times log_{10} (x/x_\text{ref})$ on both sides with $x_\text{ref} = V_\text{ref}/\sqrt{\text{Hz}}$, above equation in units of dB/$\sqrt{\textrm{Hz}}$, reads
\begin{equation}
NSD = SNR - NAB,
\end{equation}
where both, the signal-to-noise ratio ($SNR$) as well as the noise amplitude per bin ($NAB$), are given by the specifications of the ADC.
The $NAB$ is connected to the sampling rate $sr = \Delta f/2$ and given by
\begin{equation}
NAB = -10\frac{\text{dB}}{\sqrt{\text{Hz}}} \log_{10}\left(\frac{sr}{2~\textrm{Hz}}\right)  = -42 \frac{\text{dB}}{\sqrt{\text{Hz}}}.
\end{equation}
With a calculated noise spectral density ($NSD$) of $-155.5$~dB/$\sqrt{\textrm{Hz}}$, we obtain in units of spectral magnetic noise amplitude
\begin{equation}
B_{dig} = \sqrt{2} \times 10^{\left(NSD\right)/20\text{~(dB/$\sqrt{\text{Hz}}$)}} \times \frac{10}{G s} \frac{\text{V}}{\sqrt{\text{Hz}}}  = 6.2 \text{fT}/\sqrt{\text{Hz}},
\end{equation}
where we use $V_\text{ref} = 10$~V for the reference voltage of the ADC and the gain $G$ of the transimpedance amplifiers as well as the steepness $s$ recorded during the measurement.
We note, that the magnetic field noise limit due to finite ADC bit resolution is on the same order as $B_{sp}$ and $B_{sn}$.
However, such a digitization limit could easily be circumvented by digitizing the difference signal directly, allowing for a reduction of the reference voltage on the ADC or an increased gain of the transimpedance amplifiers.
Scaling the input current spectral noise density of the transimpedance amplifiers of less than 1.5~pA/$\sqrt{\text{Hz}}$ at a gain of $10^4$~V/A by the steepness $s$ (cp. Fig.~\ref{Fig:signals}), we obtain a resolution limit of $B_{amp}\sim 0.5~\mathrm{fT/\sqrt{Hz}}$ far below the other noise sources considered so far.

\subsubsection{Noise measurements}
\label{SSec:noise}
In order to \edit{assess and disentangle sources of sensor noise}, we have carried out measurements inside the MSB as shown in Fig.~\ref{Fig:setups}a.
The MSB consists of three cylindrical $\mu$-metal layers with an innermost layer diameter and length of 100~cm.
With the aid of an inner triaxial Helmholtz coil system, magnetic fields up to 100~$\mu$T can be applied in arbitrary direction.
The supply currents driving the field coils are generated by low-noise power supplies (Kepco ABC60-2DM) and low-pass filtered by custom-made LC circuits with a cut-off frequency of 0.56~Hz.
For the measurements discussed here, a magnetic field along the y-axis, perpendicular to the rotational axis of the cylindrical shields and in the plane parallel to the laboratory floor (cp. Fig.~\ref{Fig:setups}a), is used.
A field amplitude of 50~$\mu$T is generated by the use of about 0.77~A applied to the field coils.

\begin{figure}[htb]
  \includegraphics[width=8 cm]{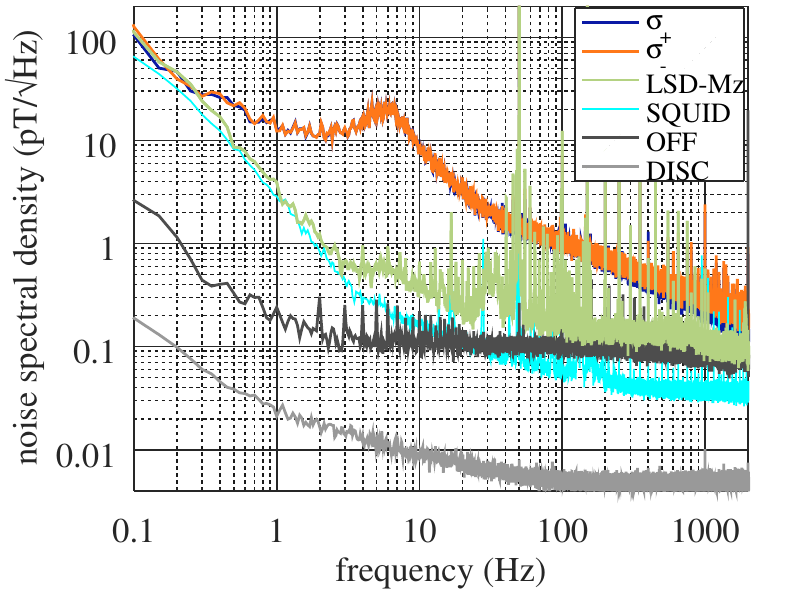}
  \caption{\edit{Noise spectral density as a function of frequency within a 50~$\mu$T field inside the MSB in magnetic field units.} The LSD-Mz signal (green) is obtained as the difference of the two single channels using $\sigma_+$ (blue)  and $\sigma_-$ (red) light, leading to an effective suppression of the common noise. As before, we have introduced a factor of 1.03 to the $\sigma_-$ signal in post processing. The cyan curve entitled SQUID represents the measured \edit{magnetic} field noise of the shielded barrel in the y-direction measured with a high temperature SQUID system operated at 77~K \cite{chwala2015}. The two gray curves OFF and DISC are measurements of the $\sigma_+$ channel when the pump laser is switched off and when cables connected to the photo diodes are disconnected from the transimpedance amplifiers, respectively. \edit{Please note that we show voltage noise spectra converted into magnetic field units using the steepness $s$ determined for the LSD-Mz difference signal in order to present them in the same figure.}}\label{Fig:noiseshield}
\end{figure}

In a first characterization step, the magnetic resonance signals are measured by sweeping the rf field frequency across the Larmor frequency.
The two photo currents as well as their difference follow a dependency similar to that plotted in Fig.~\ref{Fig:signals}.
Observation of the difference signal allows us to optimize the steepness of the LSD-Mz signal.
It is depending on the laser frequency tuning, the amplitude of the rf field, the laser power, and the cell temperature.
In the experiment, we obtain a steepness $s=3.8$~mA/$\mu$T.
Note that in post processing of the recorded data we have multiplied the signal of the $\sigma_-$ channel by a factor of 1.03.
This adjustment improves the noise characteristics as the steepness of the two channels in their crossing point is balanced.
In a next step, the rf field frequency is fixed at the zero crossing of the difference signal such that a direct reconstruction of the change of the magnetic field from the measured voltage difference of the two channels can be achieved.
\edit{Recording a time series of the voltage signal allows calculating the noise spectral density as plotted in Fig.~\ref{Fig:noiseshield} in units of magnetic field.}
Several curves are added to Fig.~\ref{Fig:noiseshield} to identify sources of noise contributing to the overall measured spectrum and to extract the basic noise level of the full device.
For better comparison and because of the illustrative character all the voltage noise spectra are scaled by the steepness $s$ to magnetic field units.
At low frequencies, below 3~Hz, the noise spectrum recorded for the LSD-Mz signal shows a 1/$f^2$ dependence and coincides with the SQUID noise spectrum (trace SQUID in Fig.~\ref{Fig:noiseshield}).
Noise measurements with the SQUID system in zero-field (field coils turned off, not shown) feature a noise level of 400~fT/$\sqrt{\text{Hz}}$ at 1 Hz, which is almost one order of magnitude below the measurements shown here, carried out at $50~\mathrm{\mu}$T.
The observed $1/f^2$ slope as well as the transition into a lower slope around 2~Hz can be accounted to the transfer function of the LC-filters used for the current driving the field coils (cut-off frequency of 0.56~Hz).
We conclude that at low frequencies the noise level is limited by the noise of the field coil current.
The requirement of magnetic field noise and stability on the order of $2\times10^{-9}$  (viz.~100~fT level in 50~$\mu$T field) poses a non-trivial technical task, especially at very low frequencies.
In our configuration, the ultimate shot-noise limit of the magnetic field generation by the discrete nature of electric charge can be estimated considering the applied coil current and the coil constant of 65.1~$\mu$T/A to be 32~fT/$\sqrt{\text{Hz}}$.

At frequencies above 3~Hz we observe the transition to a region with several distinct noise peaks and a $1/\sqrt{f}$ noise level dependence.
In additional measurements using acceleration sensors we identified the peaks between 10 and 100~Hz to result from mechanical vibrations of the MSB and the building.
We convinced ourselves that the bump at about 6~Hz visible in the spectrum of the individual channels is related to vibrations of the fiber coupling and the fiber itself.
As these are reflected as common noise on the individual channels, the bump is effectively suppressed in the LSD-Mz signal, \edit{making the system robust against movement and vibrations.}
Although, their seems to be a remaining $1/\sqrt{f}$ noise between the 6~Hz noise bump up to 100 Hz, its source could not be revealed because of the strong overlay with other noise contributions.
For frequencies above 100~Hz the \edit{noise spectral density} is dominated by the read-out electronics, as indicated by the measurement of the $\sigma_+$ channel with the pump laser switched off (trace OFF in Fig.~\ref{Fig:noiseshield}).
A noise floor of roughly 100~fT/$\sqrt{\textrm{Hz}}$ is observed in this single channel measurement, translating to about a level of 140~fT/$\sqrt{\text{Hz}}$ in the difference signal (assuming completely independent noise sources) which coincides with the noise level of the LSD-Mz signal in that frequency band.

On top of the white noise floor of the LSD-Mz as well as the SQUID signal, strong monochromatic disturbances, mostly at 50 Hz and its harmonics, are visible.
These interferences are caused by direct magnetic coupling of laboratory power lines (running at 50 Hz ac) into the MSB having a limited shielding factor in that frequency band and from electrical pick-up coupling into the magnetic field coil supply cables.
When the cables connected to the photo diodes are disconnected from the transimpedance amplifiers, the noise floor drops significantly (trace DISC in Fig.~\ref{Fig:noiseshield}), again hinting towards unresolved issues with electronic pick-up noise.

The fundamental and electronic noise sources discussed above still give room for improvement.
As we see from the measurements, at low frequencies we are limited by the characterization setup, viz. is the magnetic field generation of several tens of $\mu$T within a shielded environment.
Additionally, in the frequency range between several to 100~Hz the dominant source of noise can be related to vibrations of the characterization setup, possibly requiring additional mechanical damping techniques.
On the other hand, a higher level of integration and a rework of the electrical connection are required to reduce the pickup of high frequency noise responsible for the white noise floor of the system.

\subsection{Sensor bandwidth}
For an OPM without active feedback loop, the bandwidth is given by the width of the magnetic resonance curve.
Usually, for achieving ultimate magnetic field sensitivities, methods like SERF require narrow magnetic resonance linewidths (on the order of tens of Hz) heavily limiting the sensor bandwidth.
In contrast, in the LSD-Mz mode the linewidth is increased strongly due to the use of buffer-gas cells at elevated temperatures and due to intense optical pumping \cite{Oelsner2019}.
The LSD-Mz signal employed here has a full linewidth of about 1~kHz, resulting in a sensor bandwidth much larger than in most other OPM types.
\edit{
Additionally, the large magnetic resonance linewidth in combination with the small sensing volume ($\approx $ 4x4x4~mm$^3$ for each of the two channels), leads to an appealing behavior of the sensor under magnetic field gradients compared to OPM employing narrow resonances from large vapor cells.
As main influence in our system, a magnetic field gradient can shift the resonances of two channels constituting the LSD-Mz signal as the channels sample the local magnetic field magnitude at their position.
While the effect of a magnetic field gradient between the two sensing cavities (16 mm apart) is averaged out with respect of the zero-crossing of the LSD-Mz signal, thus not falsifying the magnetometer reading, the steepness of the LSD-Mz signal, thus the magnetometer sensitivity, may be impaired.
However, assuming a Lorentzian magnetic resonance profile with a linewidth of 1~kHz and allowing a maximal change in steepness of 5~\%, we find that magnetic field gradients on the order of 5~$\mu$T/m are tolerable.
Note that in the given setup this 5~\% change of steepness would increase the spectral noise level from 140~fT/$\sqrt{\text{Hz}}$ to 148~fT/$\sqrt{\text{Hz}}$.}

In addition to the increased sensor bandwidth resulting from the large magnetic resonance linewidth, our sensor features a locked feedback operational mode, where an active digital feedback system realized on the FPGA regulates the difference signal back to zero by controlling the rf field frequency.
This feature is not only intended to improve the sensor bandwidth but also to enable real-life applications, where the magnetic field can change very rapidly, e.g., when the sensor is moved along a magnetic field gradient or strong polarizing fields are switched on and off like, e.g., in important active geophysical exploration methods.

\begin{figure}[htb]
  \includegraphics[width=8 cm]{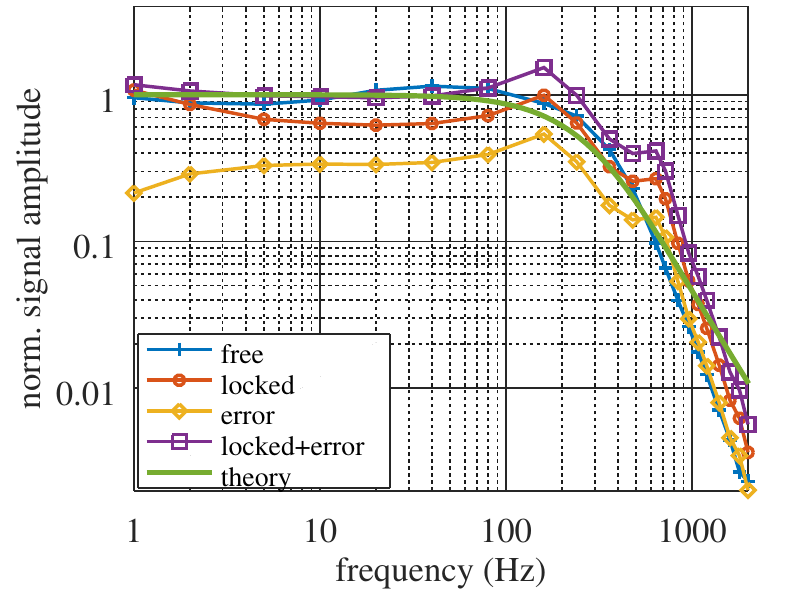}
  \caption{Normalized signal amplitude as a function of its frequency. The different curves show the LSD-Mz signal in unlocked operation (free-running) as well as the frequency feedback (locked) together with the remaining LSD-Mz signal, that in locked operation can be identified with the error signal (error). Also we included the calculated total magnetic field as sum of frequency feedback and LSD-Mz rest (locked + error) as well as a theoretical curve according to Eq.~\eqref{Eq:amp_and_phase}. }\label{Fig:bandwidth}
\end{figure}

For experimental determination of the sensor bandwidth, an additional Helmholtz coil inside the MSB is used to apply a magnetic test field $B_t$ oscillating along the pumping light propagation direction at constant amplitude.
Time series of the OPM response recorded in the free-running mode as well as with the feedback loop of the sensor closed (locked mode) at certain applied frequencies of the test field are analyzed in the frequency domain.
The accordingly extracted normalized amplitude at the frequency of the test field $B_t$ as a function of its frequency is plotted in Fig.~\ref{Fig:bandwidth}.

In the case of unlocked operation we observe a frequency response, which is flat for frequencies up to 100~Hz.
For the bandwidth a value of 250~Hz is experimentally determined, in agreement with the order of magnitude expected from the full linewidth of about 1~kHz.
\edit{In addition to the measurements,  a theoretical curve according to the difference of two channels following Eq.~\eqref{Eq:amp_and_phase} with opposing helicities has been added to Fig.~\ref{Fig:bandwidth}.
We used detunings $\delta = \pm 150$~Hz for the respective beam helicities, $1/T_1 = 200$~Hz, $1/T_2 = 350$~Hz as well as a driving amplitude $\Omega_{rf} = 10$~Hz, as reconstructed from the experiments.
The result is in good agreement to the experimental curves, however, the steepness of the frequency cut-off is still underestimated by our theory.
We attribute this to the use of a two-level approximation and neglecting the creation of higher harmonics of the signal frequency due to the nonlinearity inherent to the system response.
}

When our device is operated in the locked mode, we achieve a similar frequency response.
Depending on the control parameters set for the feedback loop, it is possible to extend the bandwidth to about 300~Hz as depicted in Fig.~\ref{Fig:bandwidth}.
\edit{However, bumps appearing at roughly 140~Hz and 600~Hz indicate an optimization potential for the digital feedback loop to allow a further enhanced sensor bandwidth.}

Note, that the minimal frequency step size in rf field generation is about 10~pT (see the discussion above), thus the feedback is not able to keep the LSD-Mz error signal strictly at zero.
We expect that this contributes to the limited bandwidth increase observed in our experiment.
Additionally, changes in magnetic field smaller than the minimal feedback step size remain on the error signal.
Therefore, as a proper output signal we consider the sum of rf frequency feedback signal and the remaining LSD-Mz signal.
In this way we reconstruct nearly the same signal amplitudes as in the free operational mode up to a signal frequency of 100~Hz.

\edit{Rescaling the noise measurements by the bandwidth measurements allows the determination of the magnetic field resolution of our device, viz. the minimal signal amplitude at a given frequency required to allow detection with the measurement system.
As the signal response amplitude drops above 250~Hz for unlocked and above 300~Hz for locked operation, the magnetic field resolution of the sensor drops for frequencies beyond the sensor bandwidth.
However, increasing the signal bandwidth by optimizing the feedback loop offers the possibility to improve the resolution limit at higher frequencies.}
\subsection{Operation in Earth's magnetic field}
In a final experiment, we compare measurements of the OPM system operated in locked mode and measurements of the commercial SQUID system already used in the MSB before (Sec.~\ref{SSec:noise}), now within Earth's magnetic field.
A photograph of the two sensor systems in place is shown in Fig.~\ref{Fig:setups}b).
The OPM, when operated within medium-sized static magnetic fields, is a scalar-type sensor, i.e. it returns the magnitude of the magnetic field in terms of the Larmor frequency of the precessing alkali atoms.
Thus, for small magnetic field changes and noise amplitudes to first order the OPM is sensitive only to field components pointing in parallel to the static magnetic field vector.
In contrast, the vectorial-type SQUID system consists of three independent magnetometers that measure the variation of the magnetic field along three mutually perpendicular Cartesian components.
Each of the three SQUID magnetometers is limited by a white noise floor of about 30~fT/$\sqrt{\text{Hz}}$.
To allow a comparison between SQUID and OPM data, we subtract the measured mean magnetic field magnitude from the OPM's time series data.

\begin{figure}[htb]
  \includegraphics[width=8 cm]{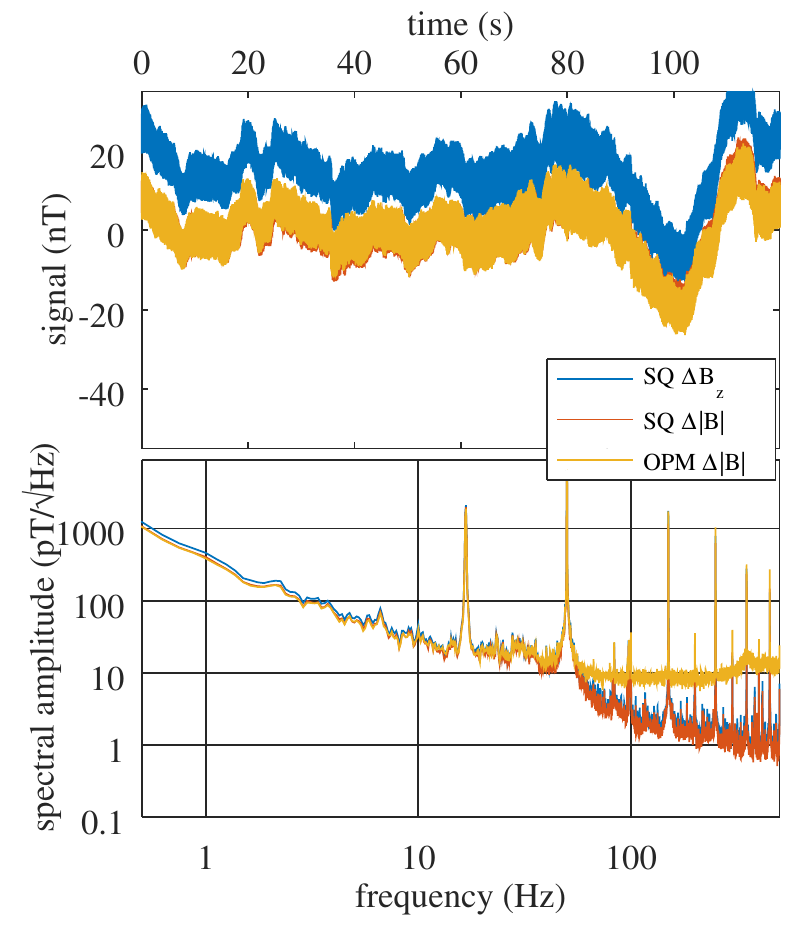}
  \caption{Time series as well as noise spectrum of the magnetic field change measured with our LSD-Mz OPM system (OPM $\Delta \left| \text{B} \right|$) and a commercial SQUID system. For the latter we plotted the z-component (SQ $\Delta \text{B}_z$) together with the recalculated absolute value change (SQ $\Delta \left| \text{B} \right|$).}\label{Fig:OPMvsSQUID}
\end{figure}

Furthermore, we placed the SQUID system such that the magnetic field vector lays inside of its x-z plane. With the Earth's magnetic field's inclination of $66^\circ31'$ the SQUID z-component thus corresponds to the OPM signal to some approximation.
In addition to the measurement of the SQUID z-component, we estimate the change in absolute magnetic field amplitude by vector addition of all three measured SQUID vector components and taking the field's inclination into account.
The measured data is shown in Fig.~\ref{Fig:OPMvsSQUID}.

The time traces recorded by OPM and SQUID follow the same trend.
The calculated cross correlation of the two data sets for a 120~s long time series is about 0.9.
Despite of the different bandwidths of the devices and the different spectral distributions of the noise having a strong influence to the cross correlation function \cite{Keshner1982}, the reduced value of the correlation can most likely be accounted to the fundamentally different measurement principles of the sensors as discussed above.
\edit{
The noise spectra measured with OPM and SQUID show an identical trend up to a frequency of about 100 Hz.
For frequencies above 100~Hz we observe a transition into a flat OPM noise limit of about 10~pT, while the amplitude of power line harmonics are in good agreement between the two measurement systems up to 450 Hz.
The flat noise level is about a factor 100 higher than in the shielded measurements.
We attribute this increase to a non-optimal balancing of the channels leading to a reduction in noise compensation, the smaller steepness of $s=$1.0~mA/$\mu$T (leading to a stronger influence of the electronic noise) as well as additional external noise coupled into the low impedance input of the transimpedance amplifiers.}

Since the noise floor is limited by the Earth's magnetic field fluctuations at very low frequencies (well below 1~Hz), in further investigations we will study ways for an efficient isolation of signals of interest in this frequency band, which is of high importance for electromagnetic methods in geophysics aiming at large exploration depths.

\section{Summary and outlook}
In summary, we have presented a application-oriented, battery-powered measurement device consisting of an LSD-Mz OPM sensor module paired with the required electronics unit.
The whole system is aimed for user-friendly operation and robustness in real world applications such as geophysical exploration.

The sensor is operated in the LSD-Mz mode and we have shown that this enables an intrinsic magnetic field noise set by the standard quantum limit at 11~fT/$\sqrt{\text{Hz}}$ and a measured white noise limit of about 140~fT/$\sqrt{\text{Hz}}$ even at Earth's magnetic field strength that is of the order of tens of $\mu$T.
We would like to stress that our sensor thus already covers more than eight orders of magnitude between absolute reading and noise limit, setting high requirements on the performance and stability of electronic feedback, rf frequency generation, as well as on the experimental characterization setup.
Our experiments show that our white noise floor is limited by electronic pick-up noise, leaving room for further technical improvements.
The low-frequency noise level is set by the surrounding magnetic field, either given by the performance of the magnetic field generation in shielded or by direct magnetic field noise in unshielded environment, respectively.
The comparison of measurements taken with the OPM and the SQUID system show high correlations of over 90~\% over a time series of 120~s, limited mainly by the fundamentally different operational modes of the two sensor types.

In future work, we will investigate the benefits of our device in various application fields, including geomagnetism, archeology, and biomagnetism.
\edit{Here, improvements of the electronic setup aimed at further noise reduction, an optimized frequency feedback, a detailed theoretical analysis of signal bandwidth limitations, as well as further improved sensor integration towards smaller system footprint and reduced power consumption are planned.}

\begin{acknowledgments}
The authors acknowledge the financial support by the Federal Ministry of Education and Research (BMBF) of Germany under Grant No. 033R130E (DESMEX). The project 2017 FE 9128, funded by the Free State of Thuringia, was co-financed by European Union funds under the European Regional Development Fund (ERDF). This work was conducted using the infrastructure supported by the Free State of Thuringia under Grant No. 2015 FGI 0008 and co-financed by European Union funds under the European Regional Development Fund (ERDF). This work has received funding from the German Research Foundation (DFG) under Grant No. SCHU 2845/2-1; AOBJ 621093.
\end{acknowledgments}
\bibliography{lit}{}
\end{document}